\documentclass[11pt,twoside]{article}

\usepackage{asp2006}
\usepackage{epsf}

\def\gla      {$\lambda$}
\def\gal      {$\alpha$}

\def\kms     {km~s$^{-1}$} 

\def\etal    {{\it et~al.}}  

\begin{document}
\title{The Structured Chromosphere and Wind of TW Hya\altaffilmark{1}}   
\author{A. K. Dupree, E. H. Avrett, N. S. Brickhouse, S. R. Cranmer,\\
\& T. Szalai\altaffilmark{2}}   
\affil{Harvard-Smithsonian Center for Astrophysics, 60 Garden Street,
Cambridge, MA 02138 USA}
\affil{}
\altaffiltext{1}{To be published in: 2007, {\it Cambridge Workshop 
on Cool Stars,
Stellar Systems and the Sun, CS14}, ASP Conf. Series, ed. Gerard van Belle.}
\altaffiltext{2}{Also University of Szeged, Hungary}    

\begin{abstract} 
The nearby accreting T Tauri star, TW Hya is viewed almost pole on, 
so its accretion and wind
characteristics can be examined.  A 
continuous set of echelle spectra of TW~Hya, 
taken with the MIKE spectrograph on the Magellan2/Clay 
telescope at Las Campanas Observatory  
in April 2006 reveals systematic variations in the flux,
velocity, and  profile of the H-alpha emission line.  These
variations appear to be consistent with the photometric period
of 2.8 days.  Absorption features recur at high outflow velocities 
(100 and 200 km s$^{-1}$) in the wind. Additional spectra from 2004
show a similar repetitive pattern.  This behavior suggests that:
(a) accretion is not uniformly distributed over the stellar hemisphere
in view; (b) stable structures are present in the chromosphere, 
most likely due to the stellar magnetic field configuration.
Semi-empirical models of the atmosphere have been constructed
to reproduce line profiles of H-\gal\ and He I, \gla10830 
using the PANDORA code and to define the
wind structure.  These preliminary  
calculations suggest the mass loss rate is
variable and comparable to H-\gal\ mass accretion rates in the 
literature, requiring a very efficient mechanism if the wind
is powered only by accretion.

\vspace*{-0.2in} 

\end{abstract}

\section{Introduction}

\begin{figure}[!ht]

\plotone{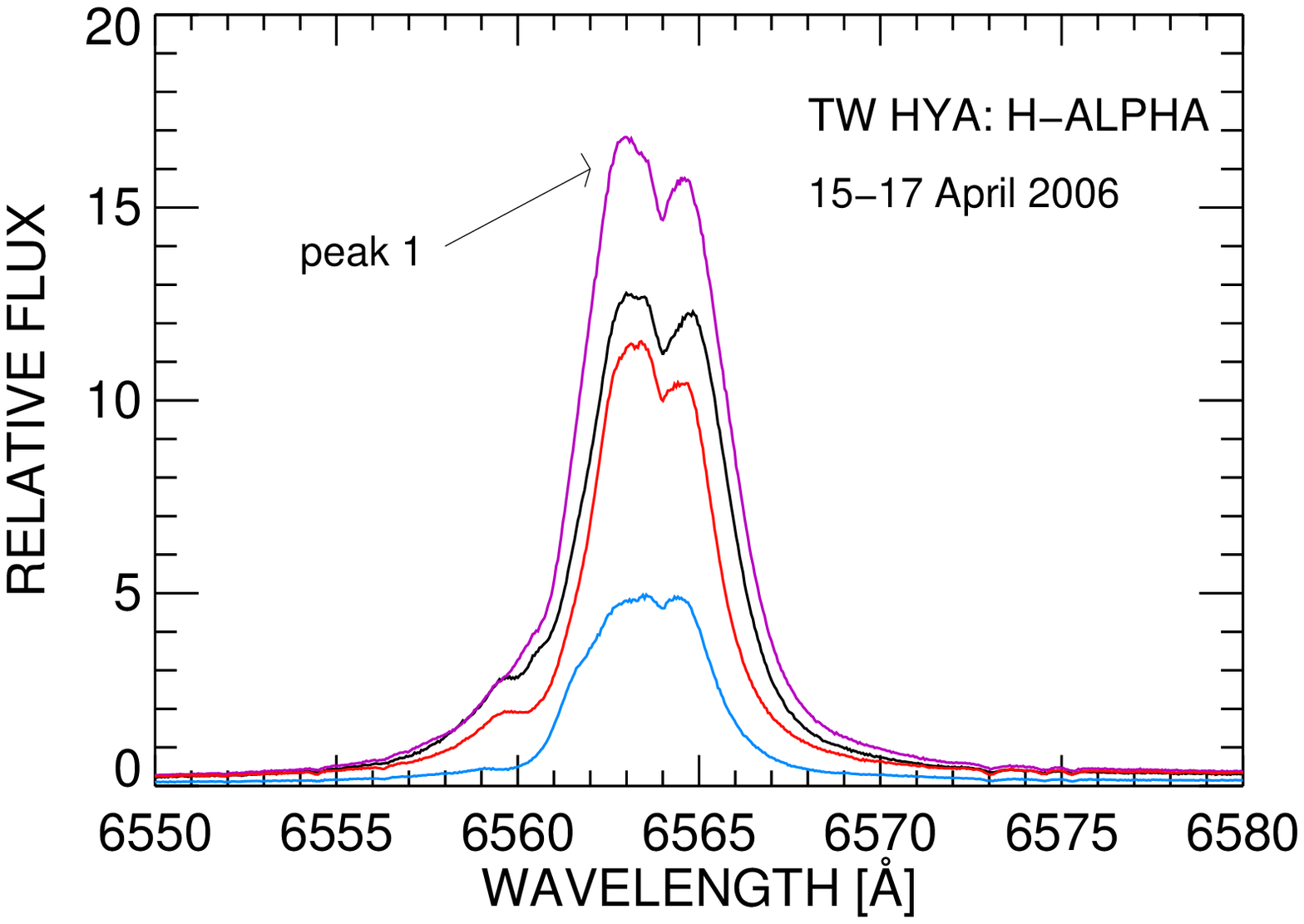}

\vspace*{-0.2in}

\caption{Sample H-alpha 
profiles of TW Hya obtained at Magellan in April 2006. The peak (Peak 1)
used to measure the  velocity
shown in Fig. 2 ({\it right panel}) is
marked. }

\plottwo{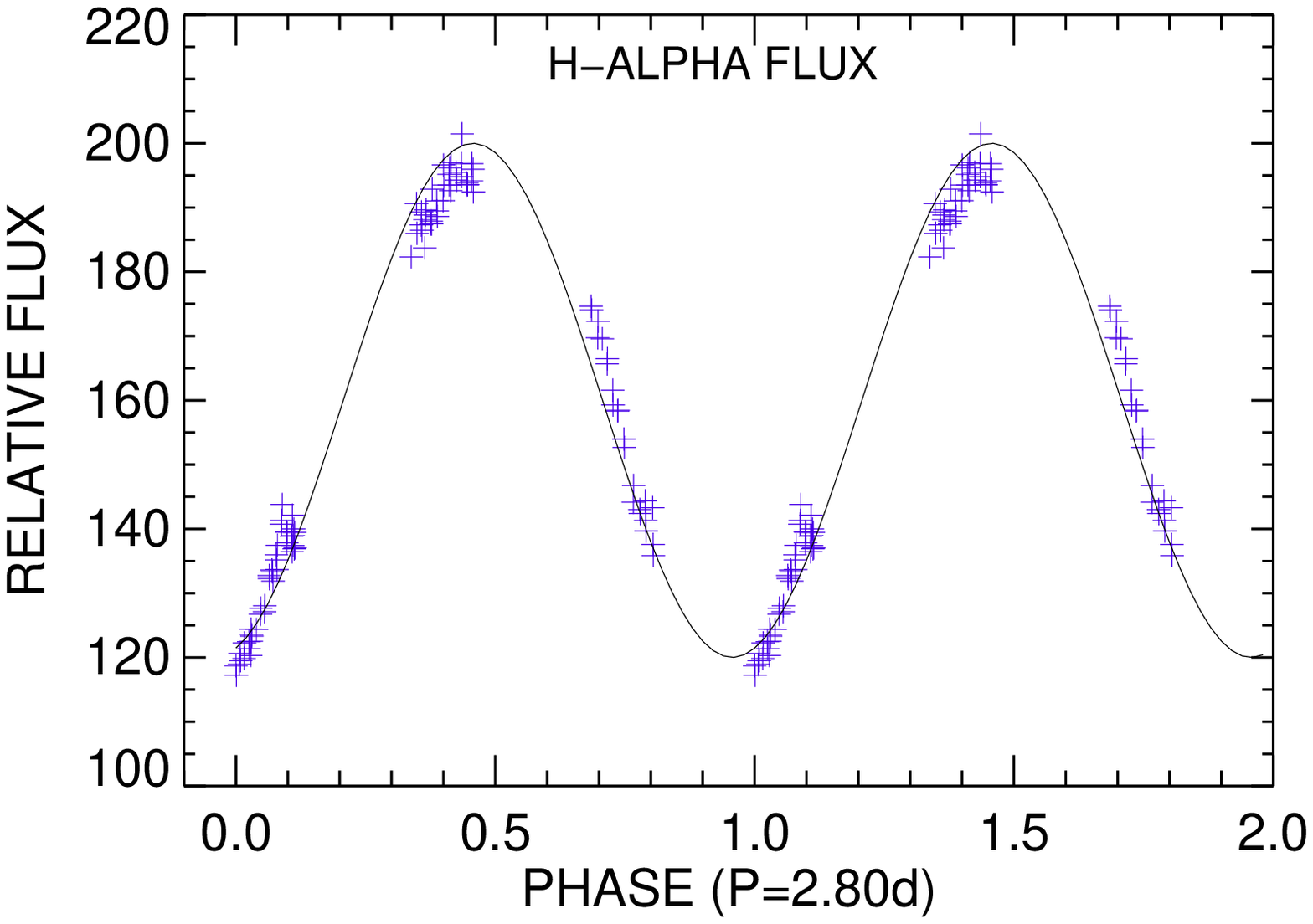}{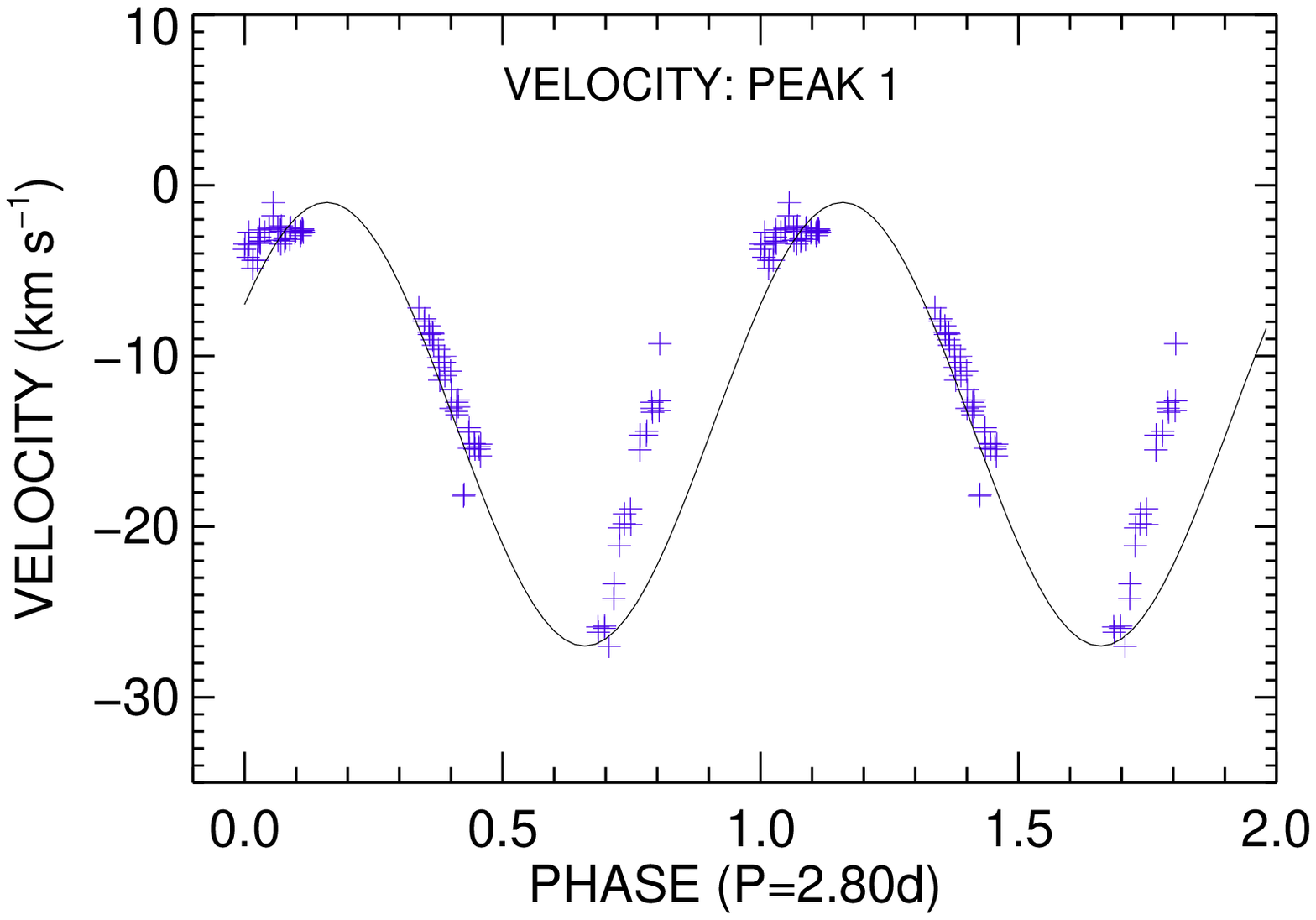}
\caption{{\it Left:} Flux in  H-\gal\  as
a function of the assumed photometric phase, Period=2.80 d (Lawson \&\
Crause 2005). Zero phase is arbitrarily set at the beginning of the 
observations.  The values are plotted twice.  A sine function has been placed on the
data to guide the eye. {\it Right:} Velocity of `peak 1'
in the H-\gal\ profile relative to the photosphere as a function of photometric
phase. A sine function is overplotted. }

\end{figure}

The nearby accreting T Tauri star, TW Hya is oriented with its
rotation axis almost along our line of sight, and the surrounding
accretion disk approximately in the plane of the sky (Krist \etal\ 2000)
providing a good opportunity to study its accretion and wind
characteristics. The discovery of a hot, fast wind from this star 
(Dupree \etal\ 2005) calls for a determination of the mass loss rate,
wind speed, and
temperature in the outer atmosphere, and its 
relation to the mass accretion rate.  A sufficiently robust 
stellar wind may
lead to optical jets, could remove angular momentum from the star  
(Matt \& Pudritz 2005), contribute to the opacity needed 
for X-ray absorption (Flaccomio \etal\ 2003), and influence the 
diminution of dust in accretion
disks (Alexander \etal\ 2005). The chromospheric lines of H-\gal\ 
and He~I \gla 10830 are reported here for TW Hya to characterize the flux
variations and line profiles.  Detailed models of wind-sensitive line
profiles of H-\gal\ and He I (\gla10830) are presented for expanding
semi-empirical atmospheric models using the 
Avrett/Loeser PANDORA code for full non-LTE
effects.  These models constrain both the atmospheric structure and
the mass loss rate implied by the observed wind-scattered 
line profiles.

\begin{figure}[!ht]

\plotfiddle{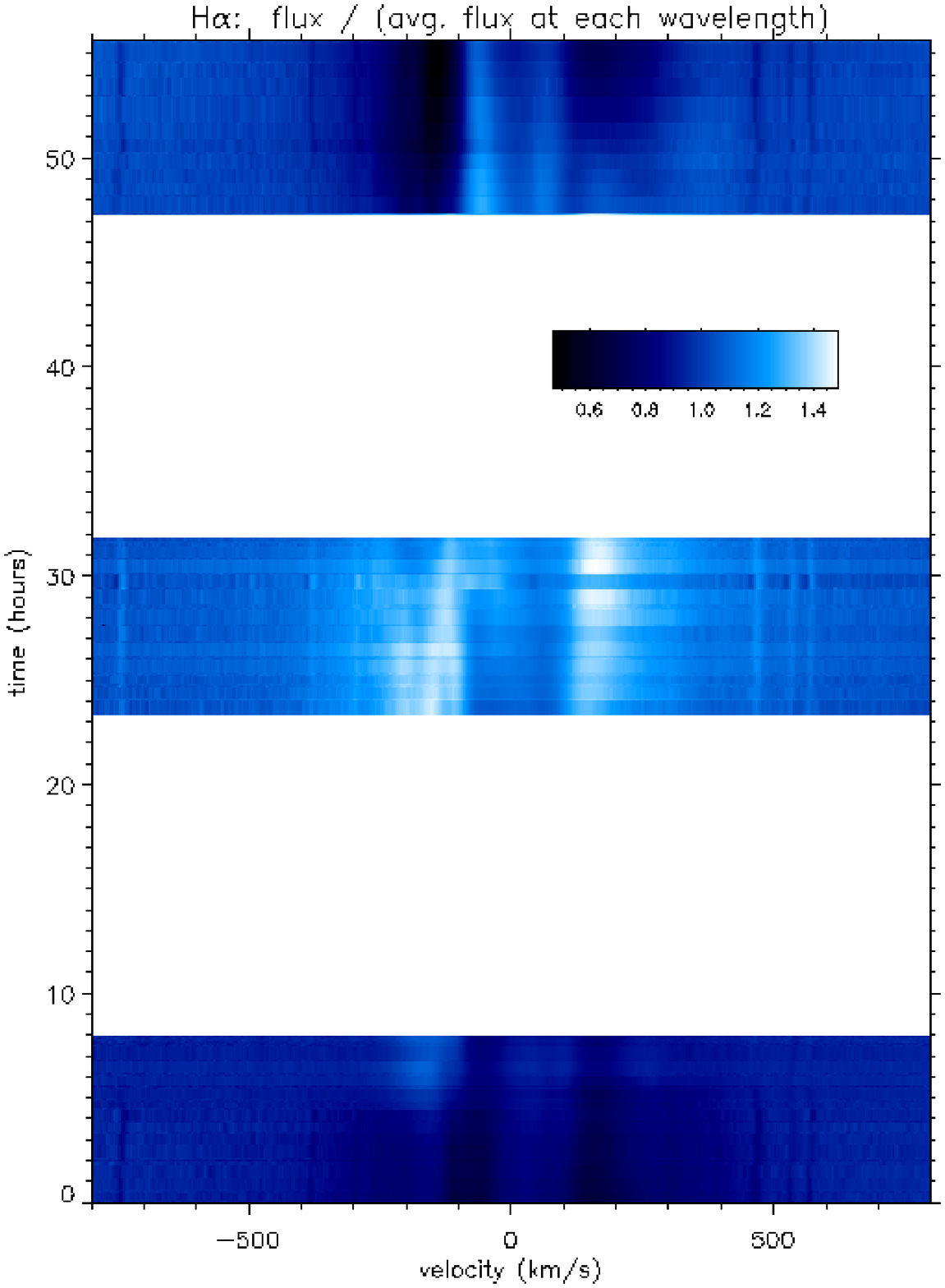}{2.5in}{0}{37}{37}{-220}{-72}

\vspace*{-3in}

\plotfiddle{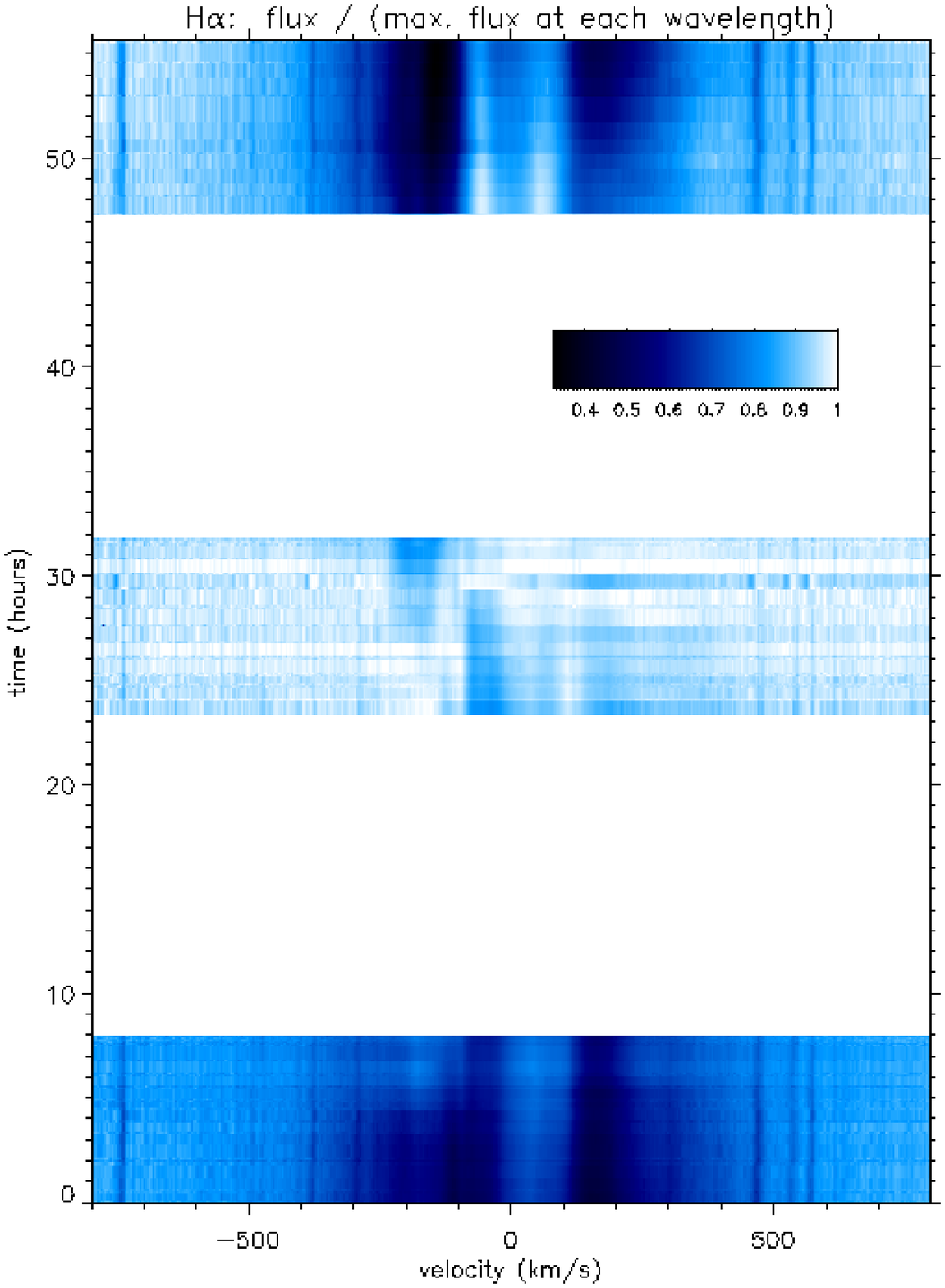}{2.5in}{0}{37}{37}{-20}{-95}

\vspace*{1.2in}

\caption{{\it Left:} Gray scale representation of the H\gal\ profiles
of TW Hya for 3 nights of observation in April 2006  
where each spectrum has been divided by the {\it average} flux at each
wavelength. The maximum H\gal\ flux that occurred during
night 2 is evident.  {\it Right:} Gray scale representation of the H\gal\ profiles
of TW Hya where each spectrum has been divided by the {\it maximum} flux
at each wavelength to display the absorption.  Absorption  at
$-$100 \kms\ 
is replaced by a discontinuous jump to a 
new absorption feature at $-$200 \kms\ during Night 2 ($\sim$29 hrs).  
These absorptions are stable in velocity for hours, 
and not similar to the moving Discrete Absorption Components (DACs)
observed in the winds of hot stars (cf. Howarth \etal\ 1995).}
\end{figure}

\section{Observations}
TW Hya was observed continuously for 3 successive nights in April 2006 at the
Magellan/Clay
telescope using the MIKE echelle spectrograph.  Additional optical spectra
were taken previously in April 2004.  The slit width of 0.75 arcsec 
yielded 
a resolution of 36,000 near H-\gal.  Infrared spectra of He~I 
\gla10830 were taken at KECK~II using the NIRSPEC infrared spectrograph
in May 2002 and July 2005, and reported elsewhere (Dupree \etal\ 2005).
Profiles of H-\gal\ showed substantial variation in flux and radial
velocity over the course of all observations, (see Fig. 1, 2, and 3).
The P Cygni profiles of He \gla 10830 showed variations in the wind
absorption among the 3 observations suggesting changes in the wind
structure.  The extent of absorption reaches $\sim\  -$300 \kms\ in both
H\gal\ and He I.
Since H-\gal\ and $\lambda$10830  are chromospheric lines, 
these velocities are supersonic and may be 
an indication of shocks and transient events.  The photospheric 
escape velocity is $\sim$500 \kms, but at a distance of 1R$_\star$
above the surface, the escape speed approaches 300 \kms\ so a small
extension of the atmosphere could easily lead to mass loss.

\begin{figure}

\plotfiddle{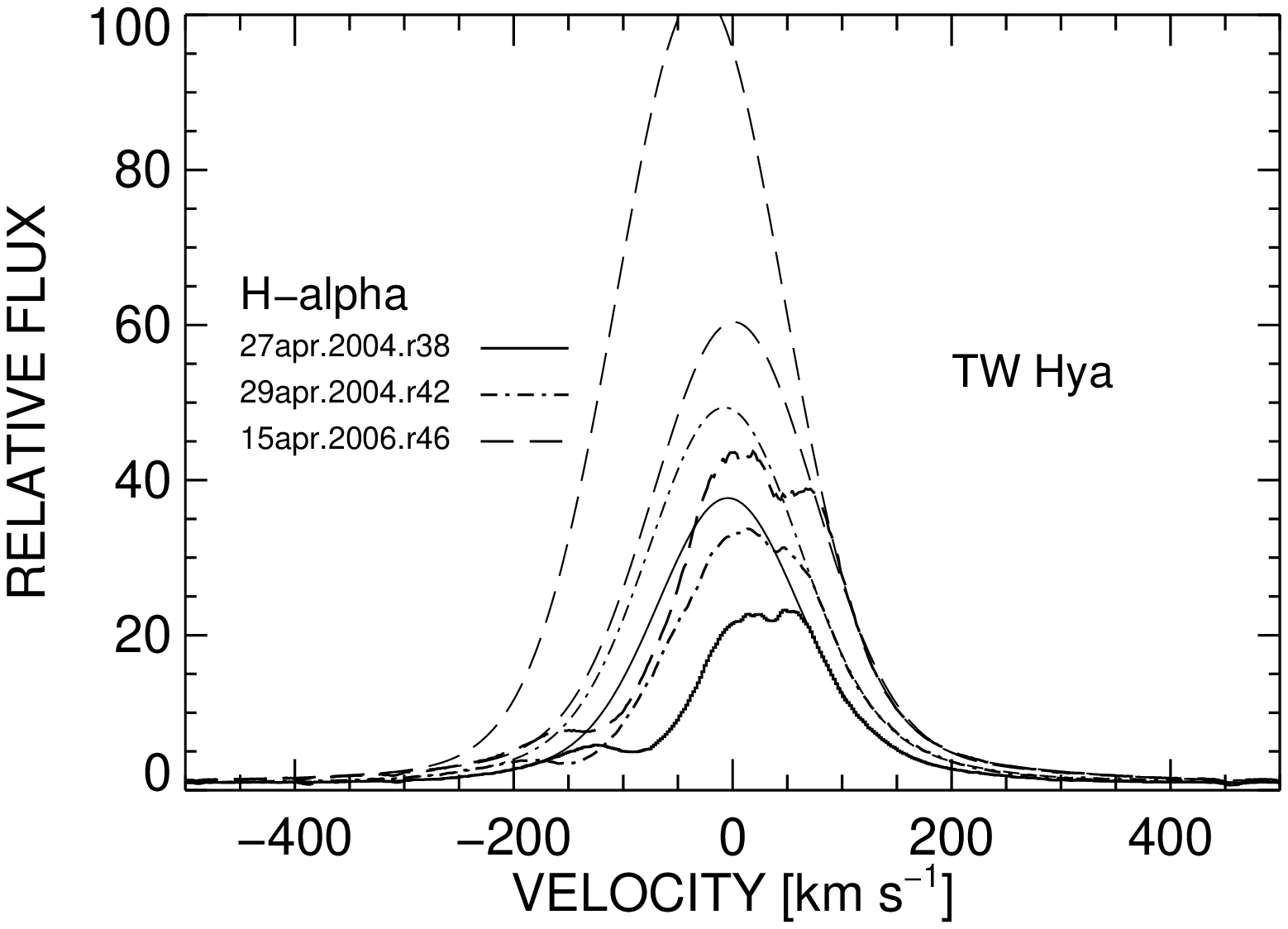}{2.5in}{0}{45}{45}{-220}{-130}
\plotfiddle{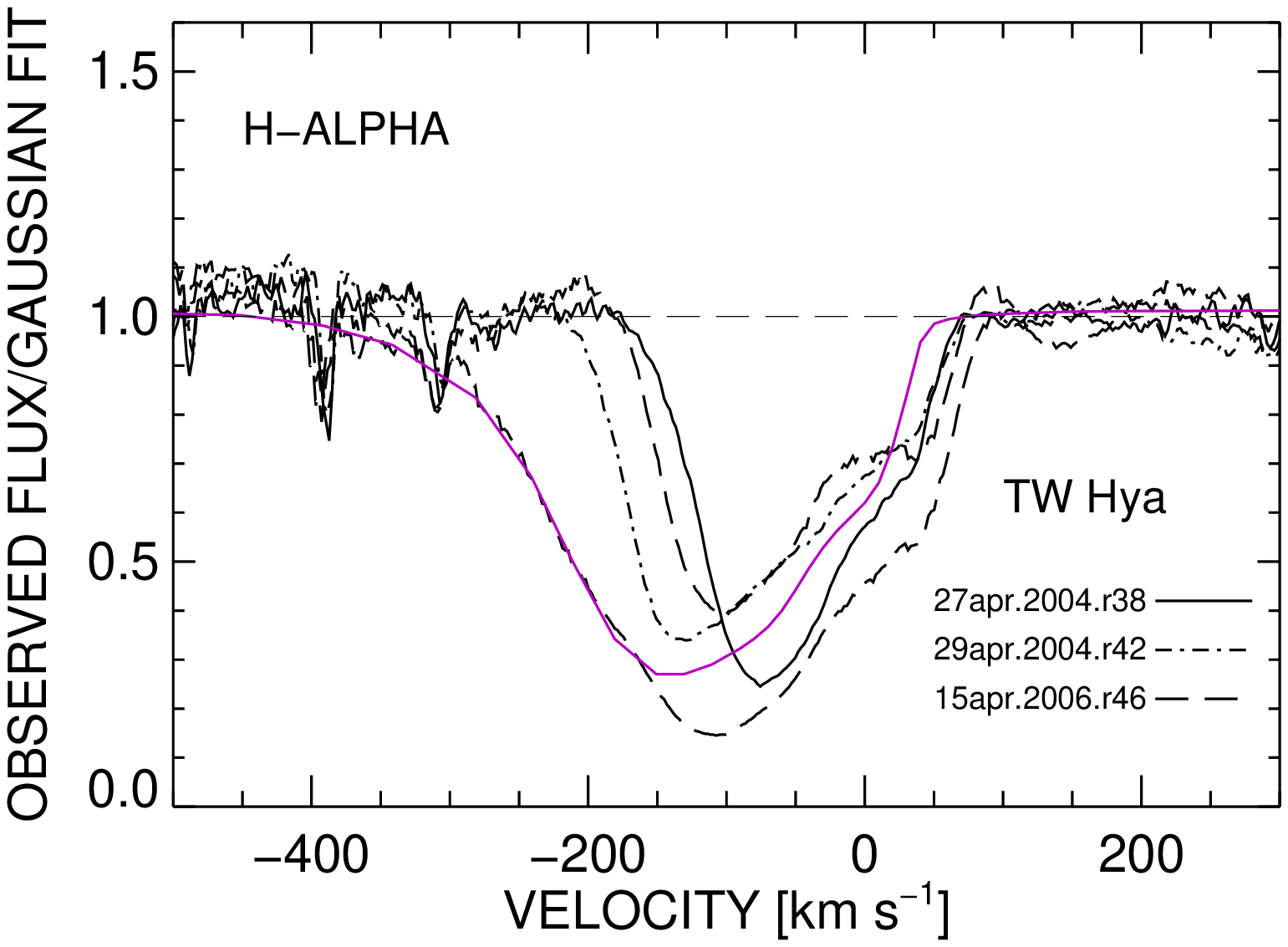}{2.5in}{0}{45}{45}{0}{+190}

\vspace*{-3.3in}

\caption{Gaussian fits to the observed H-\gal\ 
profiles on the positive velocity side ({\it left panel}) reveal 
wind absorption for comparison to a model
wind profile shown by the colored curve ({\it right panel}). Time
dependent changes in the absorption indicate that multiple models
are required with varying opacity and velocity structures.
}

\plotfiddle{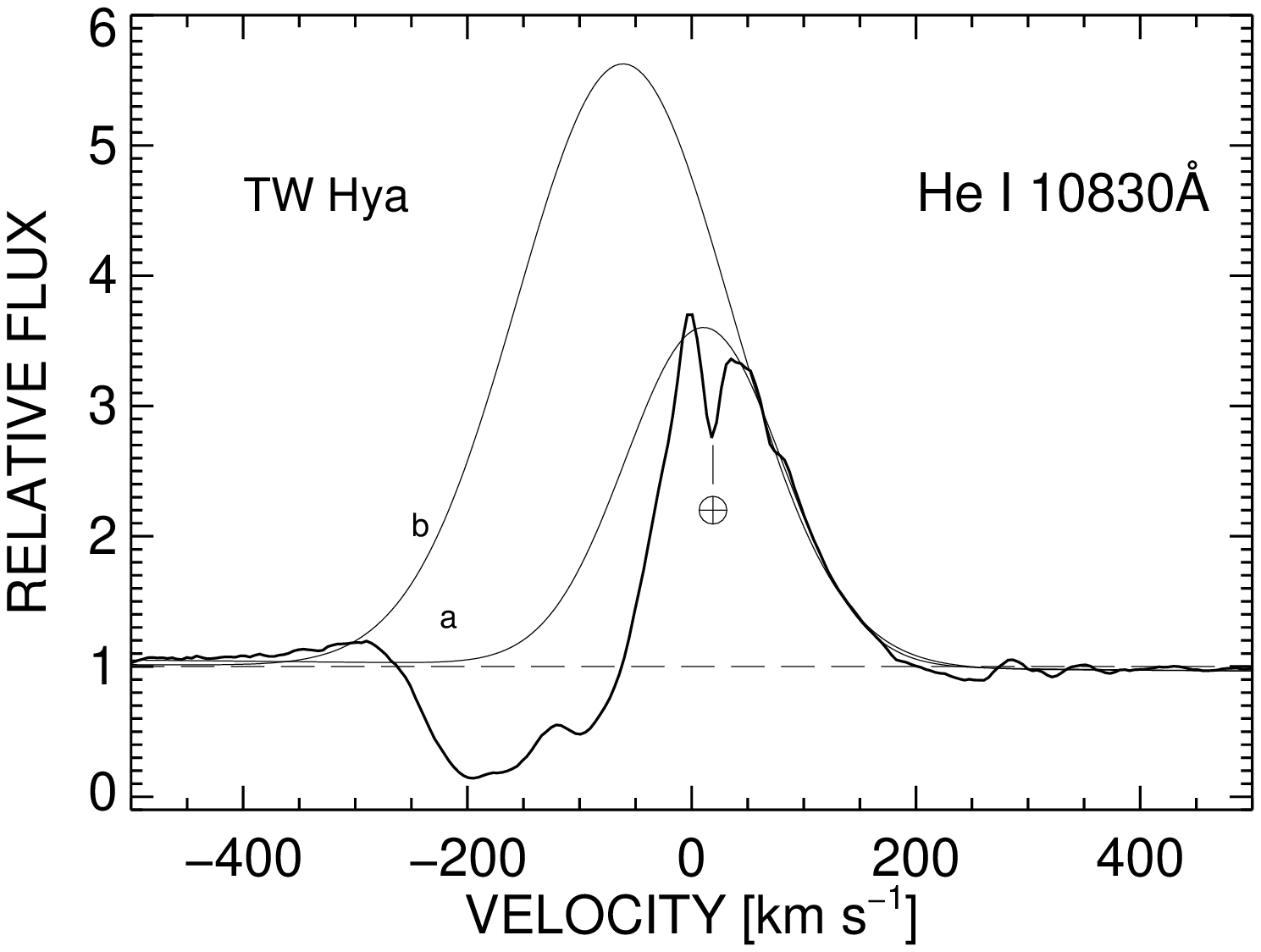}{2.5in}{0}{45}{45}{-200}{30}
\plotfiddle{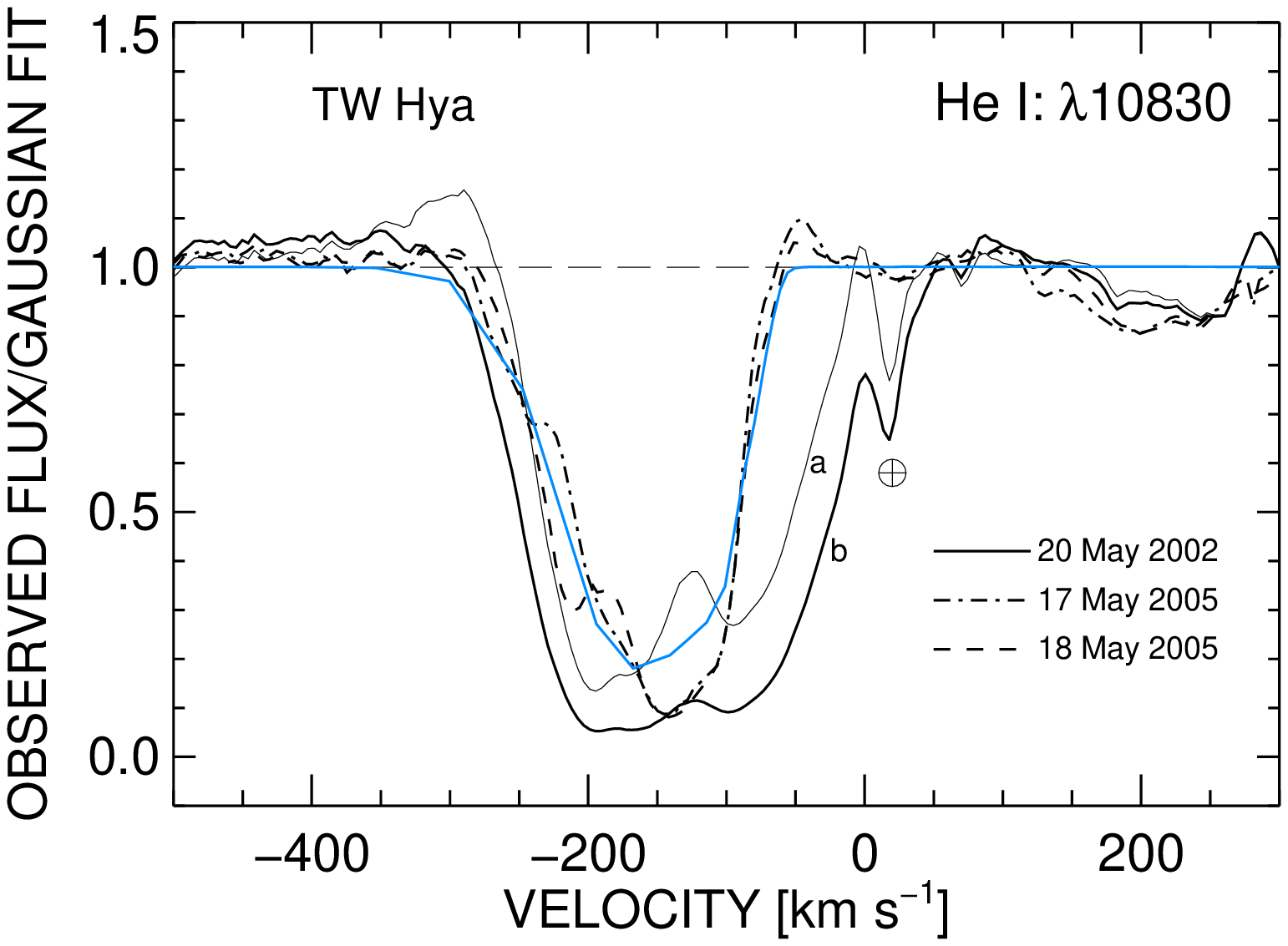}{2.5in}{0}{45}{45}{25}{+225}

\vspace*{-3.2in}

\caption{Gaussian fits to the He, \gla10830  
profile on the positive velocity side of a typical profile from 20 May
2002 ({\it left panel}) reveal 
wind absorption for comparison to the model
wind profiles shown by the colored curve ({\it right panel}).
As in the H-\gal\ line, the amount of the absorption varies,
especially
at low velocities, probably because of accretion.  
The stellar models for H-\gal\ and He I must be different which is
not unexpected since the observations are not simultaneous.}
\end{figure}

\begin{figure}

\vspace*{0.3in}

\plotfiddle{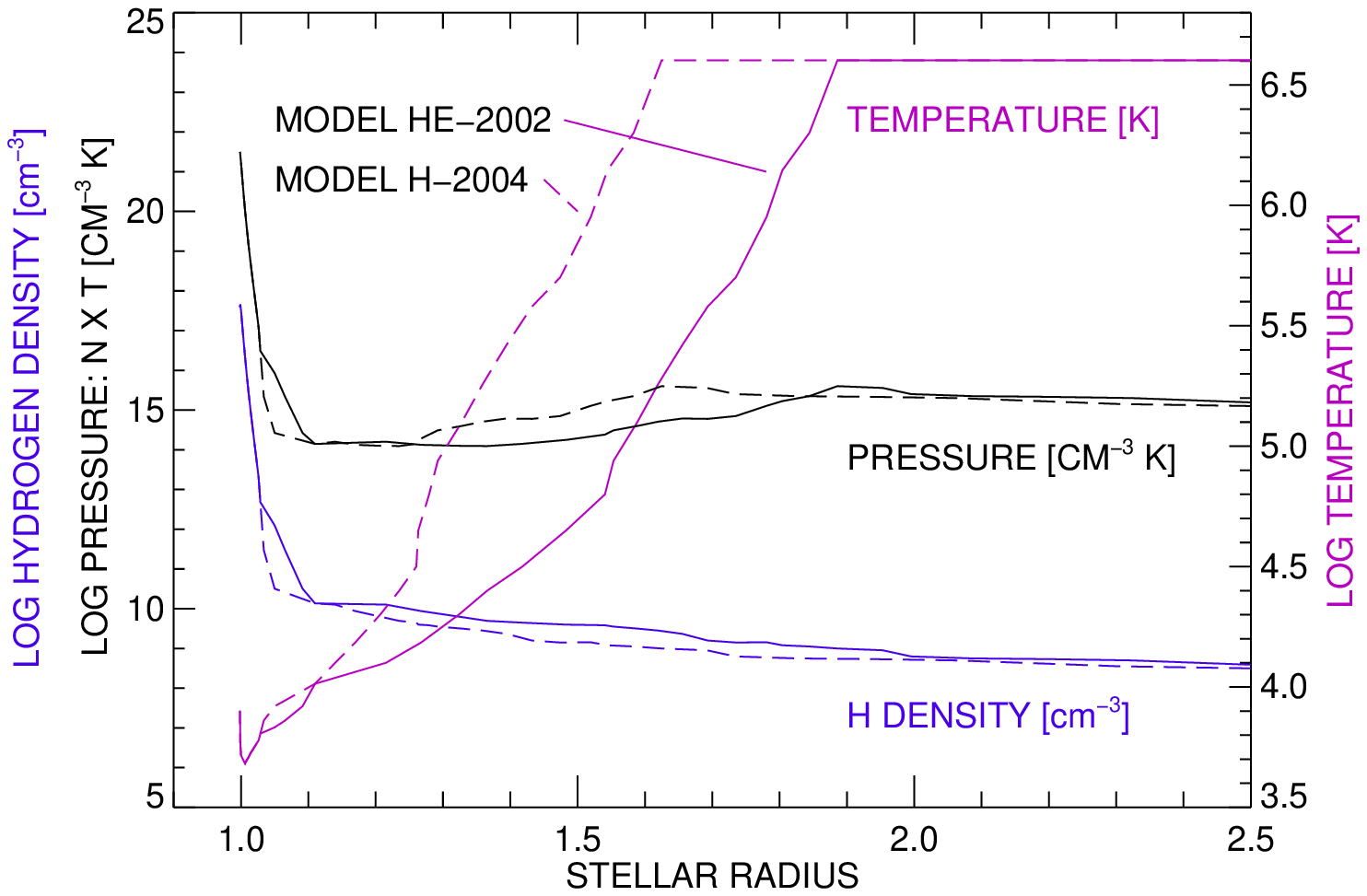}{2.in}{0}{65}{65}{-210}{-80}

\plotfiddle{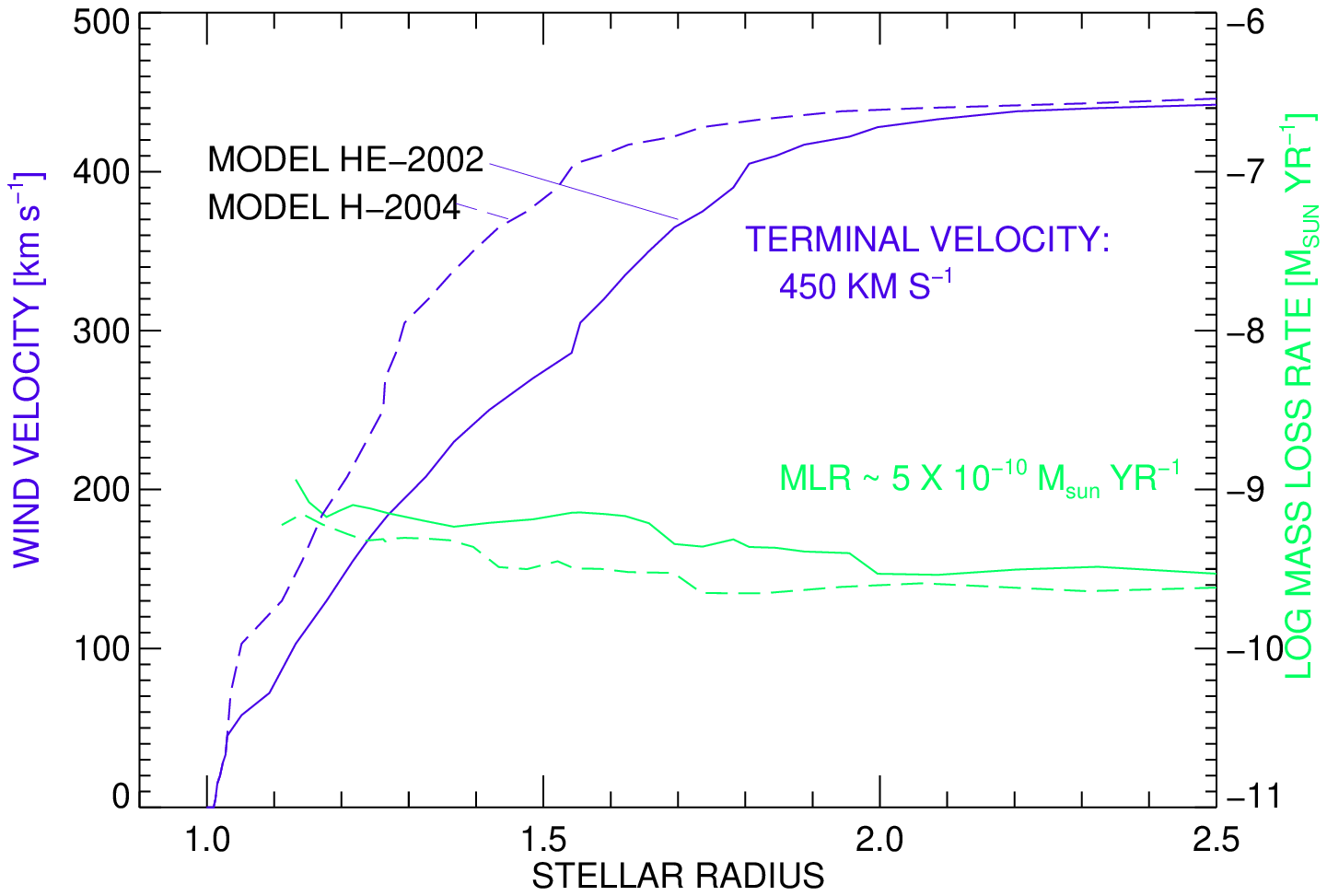}{2.in}{0}{65}{65}{-210}{-120}
\caption{Parameters for the Hydrogen Model (MODEL H-2004) and
the Helium Model (MODEL HE-2002).  The temperature variation
with radial height and the chromospheric acceleration are the principal
differences between the two models.  The mass loss rate remains
similar for both models.}

\end{figure}

\section{Models}
The PANDORA code is a general purpose non-LTE atmospheric modeling
and spectrum synthesis code (Avrett \& Loeser 2003). We invoke a model
atmosphere which is spherical and expanding. PANDORA
takes into account the time-independent optically-thick non-LTE
transfer of line and continuum radiation for
multilevel atoms and multiple stages of ionization including 
partial frequency redistribution.  In calculating 
the line source functions we include the effects of atmospheric outflows.
Multi-level atoms (3 levels for hydrogen, and
5 levels for helium, plus the continua) were used.

The starting semi-empirical models for the observations 
consist of temperature, density, and
velocity and their radial dependence (see Fig. 6).  
In this calculation, the run of
temperature and density are comparable  to solar values; 
the density at 30,000K is consistent with values from the 
diagnostic C II lines (2326\AA) in TW Hya
that indicate N$_e$ $\sim$ 4$\times$10$^{9}$ cm$^{-3}$.  The maximum coronal temperature
is taken as the value ($\sim 10^{6.6}$~K) suggested by the emission measure distribution
derived from CHANDRA X-ray spectra (Kastner \etal\ 2002). The velocity
profile is constructed to match the velocities observed in the
lines and is close to mass-conserving. The velocities and the radial height
scales are varied in order to fit the observed absorption line profiles.

The strengths of emission lines  observed in T Tauri stars are substantially influenced
by  the hot plasma produced on the stellar surface as a result of the
accretion flow.  Thus we have focussed, not on reproducing the
emission flux profile which is difficult to predict, 
but rather on the effects of the wind scattering on the
line profile. 

\section{Conclusions}
A continuous set of echelle spectra of TW Hya, 
taken with MIKE on Magellan2 in April 2006 reveals 
systematic variations in the flux, velocity, 
and the profile of the H-\gal\ emission line.  
Absorption features recur at high velocity 
in the wind. Additional spectra from 2004 show a similar repetitive pattern. 
The flux variation indicates that accretion is not uniform and
is consistent with the photometric variation found by Lawson and Crause
(2005) which
they attribute to the presence of accretion hot spots on the stellar
surface. 
Absorption `notches' appear on the short wavelength side of the
H-\gal\ line at velocities near $-$100 and $-$200 \kms.  Similar
notches
appear in the He~I \gla10830  profiles too.
The fact that similar H-\gal\ profiles (with notches) occur in our data from
2004 and 2006 indicates stable atmospheric structures.  An obvious
source creating such structures is a magnetic field that configures
the chromosphere.
Semi-empirical models of the atmosphere have been
constructed to compute the scattering line profiles using the PANDORA code. 
The wind absorption appears to vary in both H-\gal\ and He 
\gla10830, and will require multiple models.  The first approximations
shown here for H-\gal\ and He  suggest 
that the (spherical) mass outflow rate is
on the order of 5 $\times$ 10$^{-10}$ M$_\odot$ yr$^{-1}$. This
value is comparable to the accretion rate indicated by
H-$\alpha$ (Muzerolle \etal\  2000) of 
4$\times$10$^{-10}$ M$_\odot$ yr$^{-1}$ and would appear to 
require a very efficient mechanism if the wind is powered only 
by accretion.

\acknowledgements 
This research is supported in part by the Smithsonian Astrophysical
Observatory, and a Hungarian OTKA Grant No. T042509.


\end{document}